\documentclass[10pt,prl,aps,twocolumn,floatfix,showpacs]{revtex4}
\usepackage{graphicx}
\usepackage{amssymb}
\usepackage{bm}
\newcommand{\dn}{\downarrow}
\newcommand{\up}{\uparrow}

\begin{document}

\title{Ground state projection of quantum spin systems in the 
valence bond basis}

\author{Anders W. Sandvik} 
\affiliation{Department of Physics, Boston University, 
590 Commonwealth Avenue, Boston, Massachusetts 02215}

\date{\today}

\begin{abstract}
A Monte Carlo method for quantum spin systems is formulated in the 
basis of valence bond (singlet pair) states. The non-orthogonality of 
this basis allows for an efficient importance-sampled projection of the 
ground state out of an arbitrary state. The method provides access to 
resonating valence-bond physics, enables a direct improved estimator 
for the singlet-triplet gap, and extends the class of models that can
be studied without negative-sign problems. As a demonstration, the 
valence bond distribution in the ground state of the 2D Heisenberg 
antiferromagnet is calculated. Generalizations of the method to 
fermion systems are also discussed.
\end{abstract}


\maketitle

Quantum spin systems play a prominent role in current condensed matter 
 research. An increasing number of materials, with a rich variety of 
lattice geometries and spin interactions, realize many different ordered 
and disordered quantum states. On the theory side, much progress has 
been made, in particular following the discovery of superconductivity 
in doped antiferromagnetic cuprates, but there are still formidable 
challenges remaining in exploring the plethora of possible ground states 
and excitations \cite{mis05}. Gaining deeper understanding of quantum spin 
physics would not only be important in the context of particular magnetic 
systems, but could also give insights of broader relevance to correlated 
quantum matter, {\it e.g}., concerning quantum phase transitions 
\cite{sac03,sen04}. 

Spin models such as the Heisenberg hamiltonian, with interactions 
$J_{ij}{\bf S}_i \cdot {\bf S}_j$ (with spin $S_i=1/2,1$, etc.), 
can be studied by a wide range of analytical and numerical methods. 
Since all techniques have limitations, comparisons of results 
obtained in different calculations have proved to be crucial. 
Numerical finite-lattice calculations can in 
principle deliver results free of approximations, but in practice 
available computational methods are restricted to certain classes of 
models. For instance, density matrix renormalization 
\cite{whi92} is essentially limited to one dimension  and 
non-approximate quantum Monte Carlo (QMC) \cite{eve93,syl02}
can be used on a large scale only when the interactions are 
non-frustrated. Even then, there are still often 
challenges in going to lattice sufficiently large for reliable extrapolation
to the infinite lattice. Developing more general and efficient 
computational methods, for frustrated as well as non-frustrated systems, 
therefore continues to be an important field of research. 

In this Letter, a QMC method for $S=1/2$ Heisenberg models is presented 
which offers several advantages relative to state of the art ground state 
simulations (taking $T \to 0$ in world-line \cite{eve93} or 
stochastic series expansion \cite{syl02} QMC with loop-cluster updates). 
The method is formulated in an over-complete and non-orthogonal 
basis, which in the simplest case is the valence-bond (VB) basis, 
in which pairs of spins form singlets \cite{pau33,hul38,and87,lia88}. 
Any singlet state can be expressed as a sum of VB states, and the ground 
state can be projected out of an arbitrary VB state by applying a high 
power of the hamiltonian $H$. Such a scheme was used by Liang \cite{lia90}, 
who started from a good trial wave function $|\Psi\rangle$ and improved 
it by sampling terms of $(-H)^n|\Psi\rangle$. However, the projection 
stage apparently did not use importance sampling \cite{liangnote}, and an 
extrapolation in $n$ had to be performed. Santoro {\it et al.}~also used 
the VB basis, employing a Green's function method which also does not 
involve importance sampling \cite{san99}. Here it will be shown how the 
non-orthogonality of the VB basis enables a fast importance sampling 
of the terms; no variational state or extrapolations are needed.
By including triplets, excited states can also be studied, 
and unpaired spins (spinons) can be introduced as well. There is thus 
direct access to degrees of freedom that are normally not available with 
QMC but are of great theoretical interest. The valence bonds and spinons 
are the actors in resonating valence-bond physics \cite{and87}, which is 
often used as starting point for simplified quantum-dimer models 
\cite{kiv87} and field-theories \cite{sac03,sen04,lev04}. 
The method should thus facilitate closer contact with modern analytical 
treatments. Moreover, the method extends the range of models that can 
be studied without negative-sign problems.

Projection of a singlet state is here considered first, and then 
the scheme is extended to a triplet. As an illustration of results that 
can be obtained, the distribution of VB lengths in the ground state of the 
2D Heisenberg model is presented. Finally, generalizations to a wider range 
of models in other related over-complete bases are discussed.

The expansion in terms of VB states of a singlet ground state of
$N$ spins ($N/2$ valence bonds) is written as
\begin{equation}
|0\rangle = \sum_k f_k |(a^k_1,b^k_1)\cdots
(a^k_{\frac{N}{2}},b^k_{\frac{N}{2}})\rangle
= \sum_k f_k |S_k\rangle ,
\label{expansion}
\end{equation}
where $(a_i^k,b_i^k)$ denotes two spins paired up in
a singlet,
\begin{equation}
(a,b)=(\up_a\dn_b - \dn_a\up_b)/\sqrt{2},
\label{singlet}
\end{equation}
{\it i.e.}, a valence bond,
and $k$ labels all bond tilings of the lattice (allowing arbitrary 
bond lengths). The notation $|S_k\rangle$ has been introduced for 
convenience. The expansion can always be made positive definite; 
any negative $f_k$  can be made positive 
by switching the indices of one singlet.

\begin{figure}
\includegraphics[height=2.25cm]{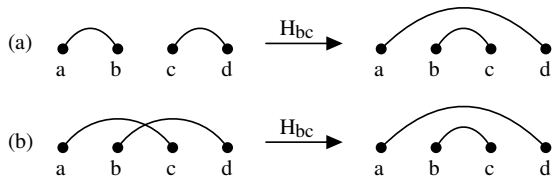}
\caption{Action of a bond operator on two VB states.}
\label{flip1}
\end{figure}

Since the VB basis is over-complete, the expansion coefficients
are not unique in general. However, the expansion of 
any state $|\Psi\rangle$ written in the VB basis of course has a unique 
expansion in energy eigenstates $|n\rangle$;
\begin{equation}
|\Psi \rangle = \sum_k g_k |S_k\rangle = \sum_n c_n |n\rangle .
\end{equation}
Therefore, acting on this state with a high power ($n\to \infty$) of the 
hamiltonian projects out the ground state:
\begin{equation}
[-(H-C)]^n |\Psi \rangle = \sum_k g_{n,k} |S_k\rangle 
\to c_0 |E_0-C|^n |0\rangle.
\label{projection}
\end{equation}
A constant $C$ has been subtracted in order to render the magnitude 
of the lowest eigenvalue larger than the highest one. QMC methods 
based on (\ref{projection}) in the standard basis of eigenstates of all 
$S^z_i$ are commonly used \cite{tri89,sor98}, although for bipartite 
systems they tend to be less efficient than low-temperature simulations 
with advanced finite-$T$ methods \cite{eve93,syl02}. However, for frustrated 
systems \cite{sor98} and t-J models \cite{hel00}, where there are 
sign problems (non-positive-definite $g_{n,k}$ and analogous mixed 
signs in other methods) projector methods are superior if a good trial
state $|\Psi\rangle$ can be used. 

\begin{figure}
\includegraphics[height=1.5cm]{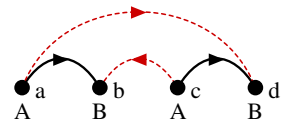}
\caption{Singlet convention for a system with sublattices A and B. 
The arrows indicate the order of the spins in the singlets, {\it e.g.}, 
$c \to b$ means $(c,b)=(\up_c\dn_b - \dn_c\up_b)/\sqrt{2}$.}
\label{flip2}
\end{figure}

The first observation underlying the projector method in the VB basis is 
that the application of a Heisenberg interaction operator on a VB state 
leads to a very simple rearrangement of valence bonds 
\cite{hul38,lia90}. Consider the Heisenberg hamiltonian, on any lattice, 
written as
\begin{equation}
H = -\sum_{\langle i,j\rangle} J_{ij}H_{ij},~~~ 
H_{ij} = -({\bf S}_i \cdot {\bf S}_j - \hbox{$\frac{1}{4}$}).
\label{heisenberg}
\end{equation}
Acting with $H_{ab}$ on a VB state in which sites $a$ and $b$
belong to the same valence bond gives an eigenvalue of unity; 
$H_{ab}|..(a,b)..\rangle=|..(a,b)..\rangle$.
Acting on sites belonging to different valence bonds gives a new 
basis state;
\begin{equation}
H_{bc}|..(a,b)..(c,d)..\rangle = 
\hbox{$\frac{1}{2}$}|..(a,d)..(c,b)..\rangle.
\label{hbcad}
\end{equation}
This {\it bond flip} is illustrated in Fig.~\ref{flip1}. Sites $a$ and $d$ 
are completely arbitrary and the sign is always positive when the indices are
in the order indicated. This implies 
that a positive-definite representation of the projection 
$(-H)^n|\Psi\rangle$ can be achieved for a bipartite lattice, by defining
$(a,b)$ so that $a$ is always on sublattice $A$ and $b$ is on sublattice 
$B$. This is illustrated using arrows on 
the bonds in Fig.~\ref{flip2}. The convention also implies a 
positive-definite expansion (\ref{expansion}) \cite{lia88}. 

For a non-frustrated interaction, one can show that
the projected ground state (\ref{projection})
contains only bonds connecting spins on different sublattices (bipartite 
bonds). Consider the VB configuration shown to the left in 
Fig.~\ref{flip1}(b). With sites a,c in sublattice A and b,d in B, both 
the bonds are non-bipartite. When the operator has acted, the new bonds 
are bipartite. With a bipartite interaction one cannot accomplish 
the reverse process (note that the hamiltonian is not manifestly hermitean 
in the VB basis) and thus, if the trial state $|\Psi\rangle$ contains 
non-bipartite bonds, they will vanish after $H$ has acted a number of 
times and cannot reappear. The ground state hence must contain only 
bipartite bonds. 

For a frustrated interaction, non-bipartite bonds are generated
and the bond flip (\ref{hbcad}) can lead to a minus sign \cite{crossnote}. 
It may also in practice not be possible to find a singlet convention 
which renders all the expansion coefficients in (\ref{expansion}) positive. 
Nevertheless, the projection scheme in principle works also for 
frustrated systems, as long as any negative signs are taken into account 
in the standard way \cite{hen00}.

With the hamiltonian (\ref{heisenberg}), $C=0$ can be used in 
(\ref{projection}). To expand $H^n$, an index sequence $P_n = [a_1,b_1],
\ldots,[a_n,b_n]$ is used to refer to an operator product 
$\prod_p H_{a_pb_p}$. One of the VB basis states can be chosen 
as the trial state;$|\Psi\rangle = |S_0\rangle$ \cite{statenote}.
The projected state is then 
\begin{eqnarray}
(-H)^n|S_0\rangle 
& = & \sum_{P_n} \prod_{p=1}^n J_{a_pb_p} 
H_{a_pb_p} |S_0 \rangle \nonumber \\
& = & \sum_{P_n} \prod_{p=1}^n w_{a_pb_p} |S(P_n) \rangle \label{proj},
\end{eqnarray}
where $|S(P_n)\rangle$ denotes the (normalized) state obtained when
the operators have acted on $|S_0\rangle$. The factors $w_{ab}$ are 
$\frac{1}{2}J_{ab}$ or $J_{ab}$ for operations with $H_{ab}$ that cause bond 
flips and are diagonal, respectively, in the course of propagating the 
state from $|S_0\rangle$ to $|S(P_n)\rangle$. The weights 
$W(P_n)=\prod_p w_{a_pb_p}$ can be identified with the expansion 
coefficients $g_{n,k}$ in (\ref{projection}). Note that no operator
$H_{ab}$ can destroy the state in (\ref{proj}), {\it i.e.}, all $W(P_n)\not=0$.
This "softness" of the basis will be taken advantage of in constructing 
an efficient importance-sampling scheme that would not be possible in 
an orthogonal ("hard") basis, where there are enormous constraints on 
the operator products.

\begin{figure}
\includegraphics[height=2.25cm]{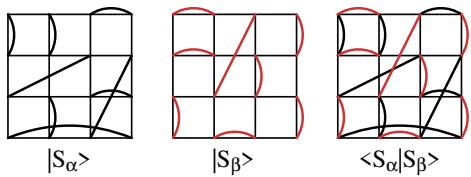}
\caption{Two VB states in two dimensions and their overlap in terms of 
loops formed by superimposing the two bond configurations. In this case 
there are $N_v=8$ valence bonds and $N_l=3$ loops, and 
$\langle S_\alpha|S_\beta\rangle = 2^{N_l-N_v} = 1/32$ \cite{sut88}.}
\label{loops}
\end{figure}

To calculate the ground state energy, the overlap with an arbitrary 
reference state $|R\rangle$ can be taken;
\begin{equation}
E_0 = \frac{\langle R|H|0\rangle}{\langle R|0\rangle} =
\frac{\sum_{P_n} W(P_n) \langle R|H|S(P_n)\rangle} 
{\sum_{P_n} W(P_n) \langle R|S(P_n)\rangle} .
\label{e0exp}
\end{equation}
One can always choose a state with equal overlap with all basis 
states, and hence all $\langle R|S\rangle$ overlaps cancel. If $P_n$ is 
sampled with probability $\propto |W(P_n)|$, the energy is
\begin{equation}
E_0  = -\frac{1}{\langle s\rangle}
\sum_{\langle i,j\rangle} J_{ij} 
\left\langle s(n_{ij} + \hbox{$\frac{1}{2}$}(1-n_{ij})q_{ij})\right \rangle,
\label{energy}
\end{equation}
where $n_{ij}=1$ (0) if there is (is not) a  bond connecting sites 
$i$ and $j$ in $|S(P_n)\rangle$. For a frustrated system, $s=\pm 1$ is 
the product of phase factors $\pm 1$ arising when propagating $|S_0\rangle$ 
to $|S(P_n)\rangle$ with the string $P_n$, and $q_{ij}=\pm 1$ arises
when $H_{ij}$ is applied once more. For a bipartite system, $s=q_{ij}=1$
and thus $E_0 = -\frac{1}{2}\sum_{ij} J_{ij}\langle n_{ij}+1\rangle$.

An expectation value 
$\langle A\rangle = \langle 0|A|0\rangle/\langle 0|0\rangle$ 
of an arbitrary operator can be 
written in terms of two projected states, obtained from the same trial 
state $|S_0\rangle$ propagated with two different operator strings
$P_n$ and $Q_n$;
\begin{equation}
\langle A\rangle = 
\frac{\sum_{P_n}\sum_{Q_n} W(P_n)W(Q_n) \langle S(Q_n)|A|S(P_n)\rangle} 
{\sum_{P_n}\sum_{Q_n} W(P_n)W(Q_n)
\langle S(Q_n)|S(P_n)\rangle}. 
\label{opexp}
\end{equation}
The weight function to be used in importance sampling is thus 
$W(P_n)W(Q_n)\langle S(Q_n)|S(P_n)\rangle$, and the operator estimator 
is $\langle S(Q_n)|A|S(P_n)\rangle/\langle S(Q_n)|S(P_n)\rangle$.

For a bipartite system, the overlap of two VB states is determined 
by the  loops formed when the bonds are superimposed \cite{sut88}, 
as illustrated in Fig.~\ref{loops} (with frustration, a sign has to 
be determined as well). Matrix elements 
$\langle S_\alpha |{\bf S}_i \cdot {\bf S}_j|S_\beta\rangle$ are also 
easily obtained from these loops; 
$\langle S_\alpha |{\bf S}_i \cdot {\bf S}_j|S_\beta\rangle/
\langle S_\alpha | S_\beta\rangle = \pm 3/4$
if sites $i$ and $j$ belong to the same loop ($+$ and $-$ for $i,j$
on the same and different sublattices, respectively, in the case of
a bipartite lattice), and $0$ otherwise.

A remarkable aspect of the VB basis is that Eqs.~(\ref{e0exp})
and (\ref{opexp}) can be efficiently sampled in an almost trivial way,
in steps where a few ($r$) of the operators in the product $P_n$ are 
changed at random. Naively one might expect that the acceptance rate 
should become very low for large expansion order $n$, but this turns out 
not to be the case. With $r=4$, the acceptance rate in the case of the 
2D Heisenberg model is $\approx 40\%$, almost independently 
of $n$ and the lattice size. The new weight can be computed by performing 
the full propagation of the state $|S_0\rangle$ with the updated product(s) 
in (\ref{proj}) [and calculating the new overlap in the case of 
(\ref{opexp})]. Recalculating the full weight, instead of just a ratio, 
may seem like an inefficient proposition. However, if $n$ has to be 
increased with the systems size as $N^\alpha$ in order to converge to
the ground state, $N^{2\alpha}$ operations [$N^{1+\alpha}$ if $\alpha < 1$ 
in the case of (\ref{opexp})] are needed to update the full operator 
sequence (attempting $n/r$ updates of $r$ operators is defined as 
one sweep; several measurements are carried out during each sweep). 
In $T \to 0$ calculations with  finite-$T$ methods 
\cite{eve93,syl02} the scaling is $N^{1+\alpha'}$ if  
$T \propto N^{-\alpha'}$. Hence, if $\alpha,\alpha' \approx 1$ the 
scaling is very similar. The gap to the lowest singlet excitation dictates 
$\alpha$ and hence in many cases $\alpha < 1$ suffices. An even faster 
sampling could likely be achieved by using a linked operator list 
\cite{syl02}; such an improvement will be left for future work.

In order to study a triplet state, consider a triplet bond;
$[a,b]_0=(\up_a\dn_b + \dn_a\up_b)/\sqrt{2}$. The eigenvalue of $H_{ab}$
operating on $[a,b]_0$ is $0$. If $H_{bc}$ is applied to $[a,b]_0(c,d)$, 
the reconfiguration of the bonds is exactly as in (\ref{hbcad}); the 
new state is $[a,d]_0(c,b)/2$. Hence, if there is no diagonal operation 
on the triplet, the triplet bond (if there is only one) behaves 
exactly as a singlet and the only change in the scheme is in the operator 
estimators. This enables an improved estimator for, {\it e.g.}, the 
singlet-triplet gap: Carrying out the simulation with only singlets, 
one of the bonds can be flagged as a triplet at the measurement stage. 
The $E_1$ estimator can be averaged over all $N/2$ initial triplet 
choices, with contributions coming only from surviving configurations, 
{\it i.e.}, those for which there are no diagonal operations on the triplet 
(the survival ratio depends on $n$). This does not change 
the scaling $N^{2\alpha}$ of the simulation and can vastly improve the 
estimate of the gap compared to $E_1 - E_0$ obtained from two independent 
simulations (the improvement is mainly due to partial 
cancelation of correlated statistical errors in $E_0$ and $E_1$). 
For example, for the 2D Heisenberg model with $N=64\times 64$, a 
projection with $n=15N$ and $10^6$ updating sweeps gave $E_0/N=0.669449(2)$ 
and the finite-size gap $E_1 - E_0 = 0.0041(2)$, corresponding to an 
accuracy gain of $60$ times for the gap, or a CPU-time reduction of $7000$. 
The energy agrees with that obtained using the SSE method \cite{syl02}; 
$E_0/N=0.669450(1)$, confirming the unbiased nature of both calculations.
 
\begin{figure}
\includegraphics[height=5cm]{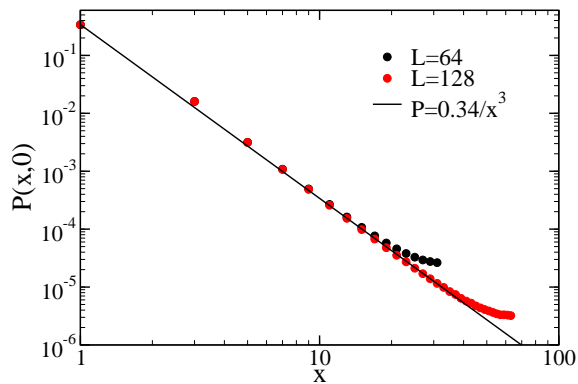}
\caption{The bond-length probability for bonds along the line $(x,0)$ 
in the 2D Heisenberg ground state wave function.}
\label{px}
\end{figure}

It is important to verify that the method works for frustrated interactions 
as well, although the basic formulation discussed here \cite{crossnote}
is not likely to be practically useful for large frustrated lattices. The
method should, however, be applicable to models with local sign problems, 
{\it e.g.}, frustrated impurities (in large host lattices). Checks against 
exact diagonalization results confirm that the scheme indeed works. For 
a $4\times 4$ system with nearest- and next-nearest-neighbor interactions 
$J_1$ and $J_2$, at $J_2/J_1 =0.1$ a ground state energy $E_0/N=0.65986(4)$ 
was obtained using $n=3N$ and $|S_0\rangle$ a columnar dimer state 
($5\times 10^9$ updating sweeps), which matches well the exact 
$E_0/N=0.659817$. The average sign in this case is $\langle s\rangle 
\approx 0.074$. 

As an application of the method, the VB length distribution in the 
ground state of the 2D Heisenberg model is presented next. 
Liang {\it et al.}~\cite{lia88} studied variational wave functions
with VB state amplitudes $f_k = \prod_i h(a^k_i,b^k_i)$ in 
Eq.~(\ref{expansion}). For $h \propto 1/r^p$, where $r$ is the bond length, 
they concluded that there is long-range N\'eel order for $p < 5$. The best 
variational energy was obtained with $p=4$, but the dependence on $p$ for 
$2 \le p \le 5$ was quite weak. The bond amplitude $h(x,y)$ does not 
correspond exactly to the probability $P(x,y) \propto \sum_k f_k n^k_{xy}$ 
[where $n^k_{xy}$ is the number of length-$(x,y)$ bonds in VB state $k$], 
but Monte Carlo simulations of the type used in 
Ref.~\cite{lia88} confirm that if $h(r) \propto 1/r^p$ then also 
$P(r) \propto 1/r^p$. Note also that $P(x,y)$ is not a ground state 
expectation value but a property of the wave function coefficients
[but the expectation value $\langle 0|n_{xy}|0\rangle$ turns out to 
be almost identical to $P(x,y)$]. A potential worry is that since 
the VB basis is over-complete, the bond distribution is not unique. However, 
the way the projection is done corresponds to a uniform averaging over 
all possible VB ground state wave functions; $P(x,y)$ defined this way 
clearly has a well-defined meaning.

Calculations were carried out on periodic $L\times L$ lattices with $L=64$ 
and $128$, with $n$ up to $15N$ and $20N$, respectively (convergence was 
checked). The results shown in  Fig.~\ref{px} suggest that $P(r) \sim 1/r^3$ 
(there is no notable angular dependence). It would be interesting to understand 
this result from an analytical starting point, and also to study the probability 
distribution for a quantum-critical system, {\it e.g.}, the Heisenberg 
bilayer \cite{san94}.

The projector scheme discussed here opens up a range of interesting
and promising avenues to be explored. The VB (+ triplets) basis is formed 
out of the 2-site eigenstates of ${\bf S}_i \cdot {\bf S}_j$ and hence is 
particularly suitable for Heisenberg models. The method can also be extended, 
without sign problems, to higher-order (non-frustrating) interactions of the 
form $-({\bf S}_i \cdot {\bf S}_j-\frac{1}{4})
({\bf S}_k \cdot {\bf S}_l-\frac{1}{4})$. This interaction
has a sign problem in the $z$-basis, and hence the present method solves 
a class of sign problems. Although there are sign problems for frustrated 
systems in general, the VB basis opens opportunities to explore cancelation 
schemes based on over-completeness \cite{crossnote}. Good 
sign-problem-free approximations could also perhaps be developed. 

It is possible to generalize the VB basis to other Hilbert spaces, 
with different types of bonds corresponding to eigenstates of the 
Hamiltonian on two sites (or even $>2$ sites, although the complexity of 
the approach then increases considerably). Such schemes for t-J and Hubbard 
models will be investigated. Although there will clearly be sign problems 
for fermions, a generalized bond-state basis also offers opportunities 
for new variational wave functions, which could be further refined with 
the projector method. Access to the VB (and generalized) degrees of freedom 
also enables construction of interesting hamiltonians acting on bonds.
Such studies could clarify the relationships between quantum dimer \cite{kiv87} 
and spin models. Studying the properties (such as the length-distribution) of 
a triplet bond in the "singlet soup" of a gapped or critical system gives 
information pertaining to the nature of the spinon bound state (magnon) or 
spinon doconfinement. This should be very useful, {\it e.g.}, in studies of 
deconfined quantum-criticality \cite{sen04,lev04}.

{\it Acknowledgments.---}
I would like to thank Kevin Beach, Subir Sachdev, and Shan-Wen Tsai for 
stimulating discussions. This work is supported by the NSF under grant 
No.~DMR-0513930.

\null


\vskip-10mm
\begin{thebibliography}{00}

\bibitem{mis05}
G. Misguich and C. Lhuillier, in {\it Frustrated spin systems}, 
edited by H. T. Diep (World-Scientific, 2005).

\bibitem{sac03}
S. Sachdev. Rev. Mod. Phys. {\bf 75}, 913 (2003).

\bibitem{sen04}
T. Senthil {\it et al.}, Science {\bf 303}, 1490 (2004).

\bibitem{whi92}
S. R. White, Phys. Rev. Lett. {\bf 69}, 2863 (1992).

\bibitem{eve93}
H. G. Evertz,  Adv. Phys. {\bf 52}, 1 (2003).

\bibitem{syl02}
A. W. Sandvik, Phys. Rev. B {\bf  59}, 14157 (1999);
O. F. Sylju{\aa}sen and A. W. Sandvik, Phys. Rev. E {\bf 66}, 046701 (2002). 

\bibitem{pau33}
L. Pauling, J. Chem. Phys. {\bf 1}, 280 (1933).

\bibitem{hul38}
L. Hulth\'en, Ark. Mat. Astron. Fys. {\bf 26A}, No.~11 (1938).

\bibitem{and87}
P. W. Anderson, Science {\bf 235}, 1196 (1987).

\bibitem{lia88}
S. Liang, B. Doucot, and P. W. Anderson, Phys. Rev. Lett. {\bf 61}, 365 (1988).

\bibitem{lia90}
S. Liang, Phys. Rev. B {\bf 42}, 6555 (1990).

\bibitem{liangnote}
The details of the procedures used in \cite{lia90} are unclear.

\bibitem{san99}
G. Santoro {\it et al.}, Phys. Rev. Lett. {\bf 83}, 3065 (1999).

\bibitem{kiv87}
S. A. Kivelson, D. S. Rokhsar, and J. P. Sethna, Phys. Rev. B {\bf 35},
8865 (1987).

\bibitem{lev04}
M. Levin and T. Senthil, Phys. Rev. B {\bf 70}, 220403 (2004).

\bibitem{tri89}
N. Trivedi and D. M. Ceperley, Phys. Rev. B {\bf 40}, 2737 (1989).

\bibitem{sor98}
S. Sorella, Phys. Rev. Lett. {\bf 80}, 4558 (1998).

\bibitem{hel00}
C. S. Hellberg and E. Manousakis, Phys. Rev. B {\bf 61}, 11787 (2000).

\bibitem{crossnote}
For frustrated models that can be written in terms of bipartite and 
non-bipartite interactions, one can rewrite 
states with non-bipartite bonds using only bipartite bonds. The projection 
can thus be restricted to the bipartite sector. Some
known singlet-product ground states, e.g., at the Majumdar-Ghosh 
point $J_2/J_1=1/2$ in the frustrated chain, are then left invariant in 
each step of the Monte Carlo projection. Whether this property could be
exploited for eliminating sign problems remains to be seen (the sign problem 
is reduced close to such points).

\bibitem{hen00}
P. Henelius and A. W. Sandvik, Phys. Rev. B {\bf 62}, 1102 (2000).

\bibitem{statenote}
The basis state $|S_0\rangle$ is arbitrary, but in order to achieve a 
faster convergence with $n$ it is often useful to take a state from a 
previous (short) projection run. Such a state is likely to have a large 
overlap with  $|0\rangle$. A variational wave function can also be 
sampled, as in \cite{lia90}.

\bibitem{sut88}
B. Sutherland, Phys. Rev. B {\bf 37}, 3786 (1988).

\bibitem{san94}
A. W. Sandvik and D. J. Scalapino, Phys. Rev. Lett. {\bf 72}, 2777 (1994).


\end{thebibliography}
\end{document}